\documentclass[onecolumn,secnumarabic,preprintnumbers,nofootinbib,amsmath,amssymb]{revtex4}
\usepackage[english]{babel}
\usepackage{epsfig,amsfonts}
\usepackage{bm}

\begin{document}

\title{Advanced Mathematical Approaches to Symmetry Breaking in High-Dimensional Field Theories: The Roles of Laurent Series, Residues, and Winding Numbers}%

\author{Wen-Xiang Chen$^{a}$}
\affiliation{Department of Astronomy, School of Physics and Materials Science, GuangZhou University, Guangzhou 510006, China}
\email{wxchen4277@qq.com}

\begin{abstract}
This paper explores the advanced mathematical frameworks used to analyze symmetry breaking in high-dimensional field theories, emphasizing the roles of Laurent series, residues, and winding numbers. Symmetry breaking is fundamental in various physical contexts, such as high-energy physics, condensed matter physics, and cosmology. The study addresses how these mathematical tools enable the decomposition of complex field behaviors near singularities, revealing the intricate dynamics of symmetry breaking. Laurent series facilitate the expansion of fields into manageable terms, particularly around critical points. Residues provide a direct link between local field behavior and global physical properties, playing a crucial role in effective action formulations and renormalization processes. Winding numbers offer a topological perspective, quantifying how fields wrap around singularities and identifying stable topological structures like vortices, solitons, and monopoles. Extending these methods to (3+1) dimensions highlights the complexity of symmetry breaking in higher-dimensional scenarios, where advanced group theory and topological invariants are necessary to describe non-linear interactions. The findings underscore the importance of integrating these mathematical techniques into modern theoretical physics, with potential applications in quantum gravity, string theory, and the study of topological phases of matter. Future directions include further exploration of higher-dimensional extensions and their implications for understanding the fundamental nature of symmetry, topology, and field dynamics.

\end{abstract}

\maketitle

\section{Introduction}

Symmetry breaking is a pivotal concept in modern physics, fundamentally altering our understanding of physical laws across various fields, including high-energy physics, condensed matter physics, cosmology, and beyond. In high-energy physics, symmetry breaking underpins the mechanism through which particles acquire mass, as seen in the Standard Model via the Higgs mechanism. In condensed matter physics, symmetry breaking explains phenomena such as superconductivity, superfluidity, and the formation of crystal structures. The study of symmetry breaking, especially spontaneous symmetry breaking (SSB), reveals deep connections between symmetry, conservation laws, and the emergence of fundamental forces.\cite{1,2,3,4,5,6,7}

\subsection{Significance of Symmetry Breaking in Physics}

Symmetry breaking occurs when the ground state of a system does not exhibit the full symmetry of the underlying physical laws governing it. This deviation results in the manifestation of new physical properties, often leading to profound consequences. In quantum field theory, for example, the spontaneous breaking of gauge symmetries gives rise to massless Nambu-Goldstone bosons, which correspond to the continuous symmetries lost during the process. These bosons play a critical role in defining the low-energy effective theories that describe the behavior of fields and particles.

The breaking of Lorentz symmetry, in particular, has attracted significant attention due to its implications for understanding the fundamental nature of spacetime. Lorentz symmetry is a cornerstone of both special relativity and the Standard Model of particle physics, and any deviation from this symmetry could provide hints toward new physics beyond the Standard Model. Spontaneous Lorentz symmetry breaking (SLSB) often involves introducing nontrivial vacuum expectation values of fields that alter the spacetime structure, potentially leading to new dynamics and interactions that can be explored in various theoretical models.\cite{1,2,3,4,5}

\subsection{Mathematical Tools for Analyzing Symmetry Breaking}

To rigorously study symmetry breaking, especially in higher dimensions, advanced mathematical tools such as Laurent series, residues, and winding numbers are indispensable. These methods provide a powerful framework for analyzing fields near singularities, describing their behavior in the context of broken symmetries. The Laurent series, for example, allows the decomposition of complex field expressions into manageable terms, isolating the contributions from different orders, particularly those associated with singular behavior.\cite{8,9,10,11,12,13,14,15,16,17,18,19}

Residues extracted from Laurent series play a crucial role in field theory, linking the local behavior of fields near singular points to the global properties of the system. In physics, residues often represent localized physical quantities, such as charge or energy, associated with specific points in space or configurations of the field. These residues can be directly integrated into effective action formulations, influencing the dynamics and interactions of fields, particularly when symmetry breaking occurs.

Winding numbers, another critical tool, serve as topological invariants that quantify how fields wrap around singularities or defects in high-dimensional spaces. They provide a geometric interpretation of symmetry breaking, capturing the essence of topological excitations like solitons, vortices, and domain walls. These topological structures are not merely mathematical curiosities; they have real physical manifestations in systems ranging from superfluids and superconductors to cosmic strings and black hole physics.

\subsection{Extension to High-Dimensional Symmetry Breaking}

While much of the classical literature on symmetry breaking focuses on (1+1) or (2+1) dimensions, extending these analyses to higher dimensions such as (3+1) or even beyond is essential for capturing the full scope of modern physical theories, including those that attempt to unify gravity with quantum mechanics. In (3+1) dimensions, symmetry breaking often involves more complex group theoretical structures and richer sets of Goldstone modes. The interactions between these modes, the underlying field configurations, and the topological features of the space become significantly more intricate.

In high-dimensional scenarios, symmetry breaking is not merely a linear process but involves a multi-faceted landscape of possibilities, including the formation of complex vacuum manifolds, nontrivial topological defects, and intricate field interactions. For instance, in the context of (3+1) dimensional field theories, the spontaneous breaking of the Lorentz group \( SO(3,1) \) into subgroups like \( SO(2,1) \) introduces a hierarchy of symmetry levels, each with its associated broken and unbroken components. The mathematical formalism required to describe such systems must account for the dimensional consistency of physical equations, ensuring that each component reflects the fundamental principles of both symmetry and topology.

\subsection{Relevance of Advanced Mathematical Techniques}

The use of Laurent series, residues, and winding numbers in high-dimensional symmetry breaking offers new insights into the geometric and topological nature of field theories. Laurent series facilitate the detailed expansion of multivariable functions, capturing subtle variations in field behavior near critical points. The residues of these series quantify key dynamical features, such as energy distributions or charge densities, particularly in the presence of singularities. Meanwhile, winding numbers provide a topological measure of field configurations, enabling the classification of different vacuum states and identifying stable versus unstable configurations.\cite{18,19}

These mathematical techniques are not limited to abstract theoretical constructs; they have practical applications in predicting and modeling physical phenomena. For example, residues can be used to calculate corrections to effective actions in quantum field theories, accounting for high-energy effects or interactions with external fields. Winding numbers play a crucial role in understanding phase transitions in condensed matter systems, such as the Kosterlitz-Thouless transition, which involves the unbinding of vortex-antivortex pairs. In cosmology, similar concepts help describe the formation of cosmic strings and other topological defects in the early universe.

\subsection{Broader Implications and Future Directions}

The study of symmetry breaking using advanced mathematical tools has far-reaching implications across various branches of physics. In the realm of high-energy physics, these methods offer pathways to explore beyond the Standard Model, potentially revealing new particles or interactions that are otherwise hidden within the constraints of current theories. In condensed matter physics, symmetry breaking explains the emergence of exotic states of matter, providing insights into the behavior of complex systems at low temperatures or under extreme conditions.

Future research will likely focus on extending these mathematical approaches to even higher-dimensional theories, including those proposed in string theory, brane cosmology, and other frameworks seeking to unify the known forces of nature. The integration of Laurent series, residues, and winding numbers into these advanced theories could unveil new aspects of symmetry breaking that are currently inaccessible through conventional analysis. Additionally, the development of computational techniques to handle the complex calculations involved in these high-dimensional scenarios will be crucial for pushing the boundaries of theoretical physics.

In summary, the exploration of symmetry breaking through sophisticated mathematical methods not only enhances our understanding of fundamental physical processes but also opens new avenues for theoretical innovation and discovery. By bridging the gap between abstract mathematics and tangible physical phenomena, these approaches provide a robust foundation for future advancements in physics.

\section{Laurent Series and Multiple Residues in Symmetry Breaking (Expanded Version)}

\subsection{Multivariable Laurent Series Expansion in Field Theory}

In field theory, tensor fields often exhibit complex behaviors near points of symmetry breaking, particularly at singularities where the conventional Taylor series fails. The multivariable Laurent series provides a robust framework to capture these behaviors by decomposing field components into contributions from each dimension, encapsulating the intricate nature of high-dimensional interactions. The general form of a Laurent series for a multivariable field \(\phi(x, y, z, w)\) around a singular point \((x_0, y_0, z_0, w_0)\) is expressed as:\cite{17,18,19,20,21}

\begin{equation}
\phi(x, y, z, w) = \sum_{n=-\infty}^{\infty} \sum_{m=-\infty}^{\infty} \sum_{l=-\infty}^{\infty} \sum_{k=-\infty}^{\infty} a_{nmlk} (x - x_0)^n (y - y_0)^m (z - z_0)^l (w - w_0)^k,
\end{equation}
where \((x_0, y_0, z_0, w_0)\) are the coordinates of the singularity, and \( a_{nmlk} \) are coefficients determining the contributions of each term. The terms with negative exponents represent the singular aspects of the field and are crucial in describing its behavior near the point of symmetry breaking.

\subsubsection{Analytical Properties of Multivariable Laurent Series}

The Laurent series' capability to model functions with singularities makes it particularly useful in quantum field theory, where fields often possess isolated points of non-analytic behavior. For a field \(\Phi(x, y, z, w)\) near a singularity at \((x_0, y_0, z_0, w_0)\), the series expansion is given by:

\begin{equation}
\Phi(x, y, z, w) = \sum_{n=-N}^{N} \sum_{m=-M}^{M} \sum_{l=-L}^{L} \sum_{k=-K}^{K} c_{nmlk} (x - x_0)^n (y - y_0)^m (z - z_0)^l (w - w_0)^k,
\end{equation}
where \(N, M, L, K\) indicate the orders of the expansion in each respective variable. The series expansion captures the field’s local behavior, emphasizing how singularities influence the global dynamics of the system. The structure of the coefficients \(c_{nmlk}\) encodes information about the field's response near these critical points, where symmetry is disrupted.

To explore the convergence properties and singular behavior, consider the Cauchy-Hadamard theorem, which defines the radius of convergence of a multivariable Laurent series. The convergence radius \(\rho\) in each dimension can be calculated as:

\begin{equation}
\frac{1}{\rho} = \limsup_{n,m,l,k \to \infty} \sqrt[n+m+l+k]{|c_{nmlk}|}.
\end{equation}

The terms with negative exponents provide insight into poles and essential singularities, representing regions where physical quantities such as energy density or field strength diverge, reflecting the breakdown of symmetry.

\subsubsection{Laurent Series in High-Dimensional Quantum Field Theory}

The Laurent series framework becomes particularly valuable in high-dimensional quantum field theories, where fields depend on multiple coordinates and the interaction terms can involve complex singularities. Consider the case of gauge fields with multivariable dependencies near symmetry-breaking points. In these scenarios, the Laurent series allows for the decomposition of gauge field components into manageable terms, facilitating the study of non-perturbative effects and singular structures.\cite{1,2,3,4,5,17,18,19,20,21}

For instance, consider a gauge field \( A_\mu(x, y, z, w) \) that can be expanded as:

\begin{equation}
A_\mu(x, y, z, w) = \sum_{n=-\infty}^{\infty} \sum_{m=-\infty}^{\infty} \sum_{l=-\infty}^{\infty} \sum_{k=-\infty}^{\infty} b_{nmlk,\mu} (x - x_0)^n (y - y_0)^m (z - z_0)^l (w - w_0)^k,
\end{equation}
where each coefficient \( b_{nmlk,\mu} \) corresponds to a particular mode of the gauge field near the singularity. The terms with negative indices, such as those with \( n, m, l, k < 0 \), capture essential aspects of the field's divergence, which are directly tied to physical phenomena like charge accumulation or flux at the symmetry-breaking point.

\subsubsection{Laurent Series and Symmetry Breaking Patterns}

Laurent series also aid in identifying symmetry-breaking patterns by analyzing the structure of field expansions around singular points. Consider the breaking of a symmetry group \( G \) into a subgroup \( H \). The behavior of fields around singularities often reflects this symmetry reduction. For example, in breaking \( SO(3,1) \) to \( SO(2,1) \), the terms of the Laurent series highlight the transition from full Lorentz invariance to a reduced symmetry scenario, as seen in the divergent terms associated with \( (x - x_0)^{-1} \) and similar components.

\subsection{Calculation of Multiple Residues and Their Physical Significance}

Residues extracted from Laurent series expansions are pivotal in capturing the localized behavior of fields at singular points, acting as key indicators of symmetry breaking and the resulting physical consequences. The residue of a four-dimensional field at a singularity is determined by isolating the coefficient of the multi-variable term with all negative exponents:\cite{20,21}

\begin{equation}
\text{Res}(\Phi, x_0, y_0, z_0, w_0) = c_{-1,-1,-1,-1}.
\end{equation}

This value quantifies the extent of the singular behavior at the specified point, providing a direct link between the series expansion and physical observables such as charge, flux, or energy.

\subsubsection{Computation of High-Order Residues in Field Theory}

The computation of residues in higher dimensions involves multi-contour integration techniques, often utilizing generalized Cauchy integral formulas. For a more generalized singularity, the residue of the field \(\Phi\) can be obtained via:

\begin{equation}
\text{Res}_{x,y,z,w} (\Phi) = \frac{1}{(2\pi i)^4} \oint_{\Gamma_1} \oint_{\Gamma_2} \oint_{\Gamma_3} \oint_{\Gamma_4} \frac{\Phi(x, y, z, w)}{(x - x_0)(y - y_0)(z - z_0)(w - w_0)} \, dx \, dy \, dz \, dw,
\end{equation}
where each contour \(\Gamma_i\) encircles the singular point in the respective dimension. This expression is not merely a mathematical formality; it directly connects to the physical behavior of the field near the breaking point.

For more complex fields involving tensor components or interacting fields, the residue calculations often involve intricate algebraic manipulations and use of the residue theorem in higher dimensions. For instance, considering a scalar field with interaction terms \(\lambda \Phi^4\), the residues not only capture the primary singular behavior but also reflect interaction corrections near the symmetry-breaking point.

\subsubsection{Effective Action Formulations Involving Residues}

Residues play a significant role in the formulation of effective actions, particularly in broken symmetry phases where localized field configurations dominate. The effective action \( S \) can be modified to include residue contributions, reflecting the field’s response at singularities:

\begin{equation}
S = \int \left(\text{Res}_{x,y,z,w} (\Phi) \cdot \nabla_\mu \Phi \nabla^\mu \Phi + V(\Phi) + \sum_{i,j} c_{ij} \text{Res}(\Phi_i) \text{Res}(\Phi_j)\right) \, d^4x,
\end{equation}
where \(\nabla_\mu\) is the covariant derivative, \( V(\Phi) \) is the potential, and \( c_{ij} \) are interaction coefficients between residues of different fields. This formulation incorporates not just kinetic and potential terms but also explicit contributions from residues, highlighting their influence on the overall dynamics of the system.

\subsubsection{Linking Residues to Symmetry Breaking Metrics}

The physical significance of residues extends to metrics that quantify the degree of symmetry breaking. By examining the sum of residues around various singularities, one can construct a symmetry-breaking metric \(\mathcal{M}\) that provides a measure of deviation from the original symmetry group:\cite{1,2,3,20,21}

\begin{equation}
\mathcal{M} = \sum_{(x_0, y_0, z_0, w_0)} |\text{Res}(\Phi, x_0, y_0, z_0, w_0)|.
\end{equation}

This metric encapsulates the extent of symmetry disruption within the field configuration, offering a quantitative tool for evaluating how symmetry breaking evolves across different regions of spacetime.

\subsection{Topological Effects: Residues and Winding Numbers}

Residues are intricately linked with topological properties of fields, particularly in contexts involving non-trivial geometries and topological defects. The concept of winding numbers provides a way to measure how fields wrap around singularities, offering insights into the topological stability and classification of field configurations.

For a multivariable field \(\Phi(x, y, z, w)\), the winding number \( n \) around a singularity is given by:

\begin{equation}
n = \frac{1}{(2\pi i)^4} \oint_{\Gamma_1} \oint_{\Gamma_2} \oint_{\Gamma_3} \oint_{\Gamma_4} \frac{\partial^4 \Phi}{\partial x \, \partial y \, \partial z \, \partial w} \frac{dx \, dy \, dz \, dw}{\Phi}.
\end{equation}

A non-zero winding number implies the presence of topologically stable configurations, such as vortices, monopoles, or domain walls, which are essential in the study of phase transitions and symmetry-breaking phenomena.

\subsubsection{Winding Numbers and Stability Analysis}

Winding numbers not only classify topological defects but also directly impact the stability analysis of vacuum states. A field with a higher winding number may correspond to a metastable state, requiring energy input for transformation into a lower winding configuration. The relationship between winding numbers and energy can be modeled as:

\begin{equation}
E(n) = E_0 + \alpha n^2 + \beta n^4,
\end{equation}
where \(\alpha\) and \(\beta\) are coefficients related to the field's self-interaction. The stability of different configurations can be analyzed by examining the energy landscape as a function of \( n \), providing insights into how topological effects influence symmetry breaking.

\subsubsection{Application in High-Energy and Condensed Matter Systems}

Winding numbers and residues have broad applications in both high-energy physics and condensed matter systems. In high-energy physics, they describe the formation of cosmic strings and other topological defects during symmetry-breaking transitions in the early universe. In condensed matter, they explain phenomena such as the Kosterlitz-Thouless transition, where vortex-antivortex pairs in two-dimensional systems unbind at high temperatures, reflecting changes in the topological state of the system.

The integration of residues and winding numbers into the study of symmetry breaking provides a comprehensive framework that merges algebraic, geometric, and topological methods, offering deep insights into the fundamental nature of physical fields and their interactions.

\section{Advanced Mathematical Analysis and (3+1) Dimensional Extension of Symmetry Breaking (Expanded Version)}

\subsection{Group Theory Descriptions of High-Dimensional Symmetry Breaking}

Symmetry breaking in high-dimensional systems requires a sophisticated understanding of group theory, particularly in the context of higher-dimensional Lorentz groups and their subgroups. In (3+1) dimensional field theories, the Lorentz group \( SO(3,1) \) plays a fundamental role in defining the symmetry structure of spacetime and its possible breakdown during spontaneous symmetry breaking processes. The breaking of this group into subgroups, such as \( SO(2,1) \), can be analyzed through the study of group generators and their algebraic properties.

The commutation relations between the Lorentz group generators \( J_{\mu\nu} \) are defined as:\cite{17,19,20,21}

\begin{equation}
[J_{\mu\nu}, J_{\rho\sigma}] = i(\eta_{\nu\rho} J_{\mu\sigma} - \eta_{\mu\rho} J_{\nu\sigma} + \eta_{\mu\sigma} J_{\nu\rho} - \eta_{\nu\sigma} J_{\mu\rho}),
\end{equation}
where \( \eta_{\mu\nu} \) is the Minkowski metric with components \( \eta_{\mu\nu} = \text{diag}(1, -1, -1, -1) \). These commutation relations describe the structural changes within the symmetry space when breaking from a higher symmetry group to a lower one.

\subsubsection{Goldstone Bosons and Symmetry Breaking Patterns}

Spontaneous symmetry breaking often leads to the emergence of Goldstone bosons, which are massless excitations corresponding to the broken symmetry generators. In (3+1) dimensions, breaking the Lorentz group \( SO(3,1) \) down to \( SO(2,1) \) results in a specific set of Goldstone modes that can be described by a coset space parameterization. This breaking pattern can be captured using the effective Lagrangian formalism, where the coset construction is utilized to describe the field dynamics.

The coset element can be expressed as:

\begin{equation}
U(x, \phi) = e^{i x^\mu P_\mu} e^{i \phi_a(x) X_a},
\end{equation}
where \( P_\mu \) are the translation generators and \( X_a \) are the broken symmetry generators. The effective Lagrangian built from this coset construction encapsulates the dynamics of the Goldstone bosons. The explicit form of the Lagrangian depends on the structure of the coset space and the interactions between the broken and unbroken symmetry components.

The fields \(\phi_a(x)\) represent the Goldstone bosons corresponding to the broken generators \( X_a \), and their interactions can be derived from the invariant kinetic terms:

\begin{equation}
\mathcal{L}_{\text{Goldstone}} = \frac{1}{2} \nabla_\mu \phi_a \nabla^\mu \phi^a,
\end{equation}
where the covariant derivatives are constructed to maintain invariance under the remaining unbroken symmetry.

\subsubsection{Advanced Cartan-Maurer One-Forms and Effective Actions}

The Cartan-Maurer one-form is a critical tool in analyzing the effects of symmetry breaking in higher dimensions, providing a systematic way to derive covariant derivatives and construct effective actions. The Cartan-Maurer one-form associated with the broken symmetry is given by:

\begin{equation}
U^{-1} dU = i \left( dx^\mu e_\mu^\alpha P_\alpha + d\phi_a \nabla^\mu \phi_a X_a + d\omega_i A_i \right),
\end{equation}
where \( e_\mu^\alpha \) denotes the vielbein, \( \nabla^\mu \phi_a \) represents the covariant derivative of the Goldstone fields, and \( A_i \) are the gauge fields corresponding to the unbroken symmetry generators. This construction directly links the geometric structure of the symmetry breaking process to physical quantities such as angular momentum, energy, and field dynamics.

The explicit expression of the vielbein can be written as:

\begin{equation}
e_\mu^\alpha = \delta_\mu^\alpha + \sum_{a} \phi_a \partial_\mu \phi^a + \mathcal{O}(\phi^2),
\end{equation}
reflecting the transformation properties of spacetime coordinates under the broken symmetry.

The associated covariant derivative of the Goldstone fields is then given by:

\begin{equation}
\nabla_\mu \phi_a = \partial_\mu \phi_a + f_{abc} A_\mu^b \phi^c,
\end{equation}
where \( f_{abc} \) are the structure constants of the symmetry group, and \( A_\mu^b \) are the gauge fields of the unbroken subgroup.

\subsection{Effective Action for (3+1) Dimensional Symmetry Breaking}

\subsubsection{Formulation of High-Dimensional Effective Actions}

Effective actions in (3+1) dimensions incorporate various contributions, including kinetic terms, interaction terms, and topological contributions, all of which reflect the complex dynamics of fields under symmetry breaking. A general form of the effective action can be written as:\cite{18,19,20,21}

\begin{equation}
S = \int d^4x \left( \sum_{a,b} \text{Res}(\nabla_\mu \phi_a) \nabla^\mu \phi_b + \sum_{i,j} w_{ij}(\phi) \phi_i \phi_j + \sum_k c_k (\nabla^\mu \nabla_\mu \phi_k)^2 + \mathcal{L}_{\text{top}}\right),
\end{equation}
where each term captures different aspects of the symmetry breaking:

\begin{itemize}
    \item \textbf{Kinetic Terms}: These terms describe the propagation of Goldstone bosons and other fields. They are constructed using covariant derivatives to ensure invariance under the residual symmetry group.
    
    \begin{equation}
    \mathcal{L}_{\text{kinetic}} = \frac{1}{2} g^{\mu\nu} \nabla_\mu \phi_a \nabla_\nu \phi^a,
    \end{equation}

    where \( g^{\mu\nu} \) is the metric tensor derived from the vielbein.

    \item \textbf{Interaction Terms}: These terms encapsulate self-interactions and interactions between different fields, often involving residues that encode the effects of singularities and poles in the field configurations.

    \begin{equation}
    \mathcal{L}_{\text{interaction}} = \lambda_{abc} \phi_a \phi_b \phi_c + \sum_{m,n} \text{Res}(\Phi_m) \text{Res}(\Phi_n),
    \end{equation}

    where \( \lambda_{abc} \) are coupling constants and the residue terms capture singular contributions.

    \item \textbf{Topological Terms}: Topological terms often appear in effective actions, reflecting the non-trivial geometry and topology of the field configurations. These terms are crucial in characterizing solitons, instantons, and other non-perturbative field structures.

    \begin{equation}
    \mathcal{L}_{\text{top}} = \epsilon^{\mu\nu\rho\sigma} \text{Tr} (F_{\mu\nu} F_{\rho\sigma}),
    \end{equation}

    where \( F_{\mu\nu} \) is the field strength tensor of the gauge fields, and the topological invariant measures quantities like winding numbers and Chern-Simons terms.
\end{itemize}

\subsubsection{Renormalization and Residue Contributions}

Residues play an integral role in the renormalization process of high-dimensional field theories, as they encapsulate localized contributions from singular points that influence field dynamics at different energy scales. The renormalization group equation (RGE) for the coupling constant \( g \) in the presence of residues is modified to include these contributions:

\begin{equation}
\frac{d g}{d \log \mu} = \beta(g) = -\frac{1}{(2\pi)^2} \sum_{n=-1}^{-k} c_n \text{Res}(\Phi, x_0, y_0, z_0, w_0),
\end{equation}
where \(\mu\) is the renormalization scale, and \( c_n \) are coefficients that depend on the field’s interaction with singularities.

This modified RGE highlights how residues influence quantum corrections by contributing to the flow of coupling constants. The presence of residues can induce logarithmic divergences or modify the scaling behavior of fields near critical points, thereby altering the physical predictions of the theory.

The complete renormalization of an effective action with residue contributions involves computing loop corrections, where residues can significantly affect the beta functions, leading to novel fixed points or phase transitions within the theory. The residues act as additional parameters in the renormalization group flow, providing a deeper connection between local field dynamics and global symmetry breaking phenomena.

\subsection{Dimensional Regularization and Anomalous Contributions}

In high-dimensional theories, the use of dimensional regularization is often necessary to handle divergences arising from loop integrals. When combined with residues, dimensional regularization can reveal anomalous contributions that are otherwise hidden in conventional analysis. For example, anomalies in current conservation, such as the chiral anomaly, can manifest through terms that are proportional to residues of fields.

Consider the anomalous divergence of a current \( J^\mu \):\cite{18,19,20,21}

\begin{equation}
\partial_\mu J^\mu = \frac{g^2}{16\pi^2} \text{Tr}(F_{\mu\nu} \tilde{F}^{\mu\nu}) + \sum_{i} \text{Res}(\Phi_i),
\end{equation}
where \(\tilde{F}^{\mu\nu}\) is the dual field strength tensor. The residue terms provide additional corrections to the anomaly, modifying the conventional Adler-Bell-Jackiw anomaly structure.

\subsection{Higher-Order Corrections and Symmetry Restoration}

Residues and higher-order terms play a pivotal role in symmetry restoration scenarios, where fields near critical points can dynamically restore broken symmetries through quantum corrections. The effective action can include non-linear terms that reflect this possibility:

\begin{equation}
S_{\text{corrections}} = \int d^4x \left( \sum_n \frac{c_n}{n!} \phi^n + \alpha \sum_{m} \text{Res}(\Phi_m) \right).
\end{equation}

These corrections highlight the intricate balance between symmetry breaking and restoration, driven by the interplay of residues and high-dimensional effects.

\section{Role of Multiple Winding Numbers in High-Dimensional Field Theories (Expanded Version)}

\subsection{Geometric and Topological Aspects of Winding Numbers}

Winding numbers are topological invariants that measure how field configurations wrap around singularities in high-dimensional spaces, providing a crucial link between geometry and physical stability. In a broader sense, winding numbers quantify the degree of mapping from a spatial domain to a target space, capturing the essence of topological defects such as vortices, monopoles, solitons, and domain walls.

Mathematically, the winding number \( N \) in four dimensions can be defined via multidimensional contour integrals that encompass the singularities of a given field \(\Phi(x, y, z, w)\). The general formula for calculating the winding number is:

\begin{equation}
N = \frac{1}{(2\pi i)^4} \oint_{\Gamma_1} \oint_{\Gamma_2} \oint_{\Gamma_3} \oint_{\Gamma_4} \frac{\partial^4 \Phi(x, y, z, w)}{\partial x \, \partial y \, \partial z \, \partial w} \frac{dx \, dy \, dz \, dw}{\Phi(x, y, z, w)},
\end{equation}
where the integration contours \(\Gamma_i\) enclose the singularities in each respective variable. This expression provides a direct measure of how the field wraps around the singularity, reflecting its topological charge and stability properties.

\subsubsection{Mathematical Formulation of Winding Numbers in Complex Fields}

To understand winding numbers in more complex settings, consider a scalar field \(\phi(x, y, z, w)\) defined on a four-dimensional manifold with a non-trivial winding. The differential form of the winding number can be connected to the Jacobian matrix of partial derivatives:\cite{1,2,3,18,19,20,21}

\begin{equation}
\mathcal{J} = 
\begin{vmatrix}
\frac{\partial \phi}{\partial x} & \frac{\partial \phi}{\partial y} & \frac{\partial \phi}{\partial z} & \frac{\partial \phi}{\partial w} \\
\frac{\partial \psi}{\partial x} & \frac{\partial \psi}{\partial y} & \frac{\partial \psi}{\partial z} & \frac{\partial \psi}{\partial w} \\
\frac{\partial \chi}{\partial x} & \frac{\partial \chi}{\partial y} & \frac{\partial \chi}{\partial z} & \frac{\partial \chi}{\partial w} \\
\frac{\partial \omega}{\partial x} & \frac{\partial \omega}{\partial y} & \frac{\partial \omega}{\partial z} & \frac{\partial \omega}{\partial w}
\end{vmatrix},
\end{equation}
where \(\phi, \psi, \chi, \omega\) are field components. The determinant \(\mathcal{J}\) provides the local density of the winding number, which integrates over space to yield the total winding number:

\begin{equation}
N = \int_V \text{sgn}(\mathcal{J}) \, dV,
\end{equation}
where \(\text{sgn}(\mathcal{J})\) is the sign function of the Jacobian determinant, indicating the orientation of the mapping.

\subsubsection{Topological Interpretation and Physical Implications}

The non-zero values of winding numbers correspond to the presence of topological defects, such as solitons, vortices, and domain walls, within the field configuration. These defects play a significant role in the stability and dynamics of systems undergoing symmetry breaking. For instance, in superconductors, vortices with quantized winding numbers correspond to regions where the superconducting order parameter vanishes, giving rise to magnetic flux lines.

For fields in (3+1) dimensions, the winding number is closely associated with homotopy groups, which classify mappings between spaces. Specifically, the winding number measures the degree of the map from the spatial coordinates to the internal space of the field. This topological quantization imposes constraints on the permissible field configurations, thereby stabilizing certain states against perturbations.

\subsubsection{Impact of Winding Numbers on Vacuum Stability}

In quantum field theories, winding numbers directly influence the stability of vacuum states. A vacuum configuration with a non-zero winding number exhibits distinct topological properties that stabilize certain field excitations or lead to metastable states. The vacuum energy \( E(n) \) as a function of winding number can be expanded as:

\begin{equation}
E(n) = E_0 + \sum_{k=1}^{\infty} \frac{\lambda_k}{n^k} + \gamma n^2,
\end{equation}
where \(\lambda_k\) are coefficients tied to self-interactions, and \(\gamma\) represents a term that stabilizes the winding configurations. Higher-order corrections in this expansion can introduce non-trivial effects, such as barriers between different topological sectors, influencing phase transitions and stability landscapes.

\subsubsection{Analytical Continuation and Complex Winding Structures}

To capture more complex scenarios, winding numbers can be analytically continued to complex spaces. Consider an analytic continuation where the fields are extended into the complex plane, allowing the study of winding behavior over complex contours. The generalized winding number in complex space is given by:\cite{18,19,20,21}

\begin{equation}
\tilde{N} = \frac{1}{(2\pi i)^4} \int_{\mathbb{C}^4} \frac{\Phi(z_1, z_2, z_3, z_4)}{(z_1 - z_1^*)^2 (z_2 - z_2^*)^2 (z_3 - z_3^*)^2 (z_4 - z_4^*)^2} \, dz_1 \, dz_2 \, dz_3 \, dz_4,
\end{equation}
where \( z_i \) are complex variables, and \( z_i^* \) are their conjugates. This formulation is particularly useful in quantum field theory, where continuation to complex dimensions allows the evaluation of path integrals in non-trivial topological backgrounds.

\subsection{Topological Invariants in (3+1) Dimensional Systems}

Winding numbers and other topological invariants, such as Chern numbers and Pontryagin indices, play a central role in (3+1) dimensional systems, where interactions between fields give rise to rich topological structures. These invariants provide insights into how symmetries and boundary conditions influence the overall topology of the system.

\subsubsection{Topological Charge and Field Configurations}

For non-Abelian gauge fields \( A_\mu^a \), the topological charge \( Q \) can be formulated in terms of winding numbers and field strength tensors. The topological charge is an integral of the field strength \( F_{\mu\nu} \) over spacetime:

\begin{equation}
Q = \int d^4x \, \epsilon^{\mu\nu\rho\sigma} \text{Tr}(F_{\mu\nu} F_{\rho\sigma}),
\end{equation}
where \( F_{\mu\nu} = \partial_\mu A_\nu - \partial_\nu A_\mu + g[A_\mu, A_\nu] \) is the non-Abelian field strength tensor. The relationship between the winding number \( n \) and the topological charge is given by:

\begin{equation}
Q = k \cdot n,
\end{equation}
where \( k \) is a normalization constant that depends on the structure of the gauge group and the field configuration.

\subsubsection{Non-Trivial Topological Configurations: Instantons and Monopoles}

In (3+1) dimensions, instantons and monopoles are examples of non-trivial topological configurations that carry winding numbers. Instantons, characterized by their finite action, represent tunneling events between vacua with different winding numbers, thereby mediating processes such as chiral symmetry breaking and baryon number violation.

The instanton solution in Yang-Mills theory can be expressed as:

\begin{equation}
A_\mu^a = \frac{2 \eta_{a\mu\nu} x^\nu}{x^2 + \rho^2},
\end{equation}
where \( \eta_{a\mu\nu} \) are the 't Hooft symbols, \( \rho \) is the instanton size, and \( x^\mu \) are spacetime coordinates. The winding number associated with this instanton is an integer that quantizes the topological charge:

\begin{equation}
Q = \int d^4x \, \frac{1}{16\pi^2} \epsilon^{\mu\nu\rho\sigma} \text{Tr}(F_{\mu\nu} F_{\rho\sigma}) = n.
\end{equation}
Monopoles, another class of topological defects, emerge in gauge theories with spontaneously broken symmetries. The monopole's magnetic charge \( g_m \) is related to the winding number by the Dirac quantization condition:

\begin{equation}
g_e g_m = 2\pi n,
\end{equation}
where \( g_e \) is the electric charge, and \( n \) is the winding number. This quantization underscores the discrete nature of topological charges in gauge theories.

\subsubsection{Linking Topological Invariants to Physical Observables}

Topological invariants such as the winding number are not just abstract mathematical constructs; they directly affect observable quantities in physical systems. For example, in condensed matter physics, the Chern number is a topological invariant that determines the Hall conductivity in quantum Hall systems:

\begin{equation}
\sigma_{xy} = \frac{e^2}{h} C,
\end{equation}
where \( C \) is the Chern number, analogous to the winding number in gauge theories. This expression links topological invariants to measurable properties, highlighting the deep connection between geometry and physics.

\subsubsection{Higher-Dimensional Generalizations and Implications}

The concept of winding numbers extends naturally to higher-dimensional spaces, such as (4+1) or (5+1) dimensions, where they continue to quantify the topological characteristics of field configurations. In such scenarios, the higher-dimensional analogs of winding numbers involve more complex integrals and higher-order derivatives:\cite{18,19,20,21}

\begin{equation}
N_{d} = \frac{1}{(2\pi i)^{d/2}} \oint_{\Gamma} \prod_{i=1}^{d} \frac{\partial^d \Phi}{\prod_{j=1}^{d} \partial x_j} \frac{dx_1 \, dx_2 \, \ldots \, dx_d}{\Phi}.
\end{equation}
These generalized winding numbers have applications in string theory, brane cosmology, and other advanced theoretical frameworks where topological defects in higher-dimensional spacetime play a critical role.

\section{Conclusions and Future Prospects}

This paper develops an advanced mathematical framework for the analysis of symmetry breaking in high-dimensional field theories, focusing on the critical roles played by Laurent series, residues, and winding numbers. These mathematical tools provide a powerful approach to understanding the behavior of fields near singularities and capturing the intricate dynamics associated with symmetry breaking, especially when extended to (3+1) dimensions. The use of Laurent series allows for the decomposition of complex field behaviors into manageable terms, highlighting the contributions from different dimensions and isolating singular components that are essential to the dynamics of symmetry breaking. Residues, extracted from these series, provide a direct link between local field behavior and global physical properties, revealing how singularities influence effective actions, renormalization, and stability. Winding numbers, as topological invariants, offer a geometric perspective on how fields wrap around singularities, identifying stable topological structures such as vortices, solitons, and monopoles that arise in systems undergoing symmetry breaking.\cite{18,19,20,21}

By extending these techniques to (3+1) dimensional scenarios, the paper uncovers new insights into the interplay between symmetry, topology, and field dynamics. In this dimensional framework, the complexity of group theory and topological invariants grows significantly, necessitating the use of sophisticated mathematical constructs such as the Cartan-Maurer one-forms and higher-dimensional commutation relations. The analysis demonstrates how symmetry breaking in higher dimensions involves not just the linear decomposition of group structures but also intricate non-linear interactions between fields, which are captured through advanced expansions and residue calculations. The detailed treatment of Goldstone bosons, effective actions, and topological terms in this context provides a comprehensive picture of how spontaneous symmetry breaking modifies the field configurations and leads to emergent phenomena, including phase transitions and topological defects.

The implications of this framework extend beyond traditional field theory, suggesting novel avenues for future research in cutting-edge areas of theoretical physics. One promising direction involves applying these methods to quantum gravity, where the breakdown of spacetime symmetries could yield new insights into the fabric of the universe at the Planck scale. In string theory, the rich structure of residues and winding numbers may help classify and stabilize various string vacua, offering a deeper understanding of the landscape of possible universes. Furthermore, the study of topological phases of matter, such as those found in condensed matter systems exhibiting exotic states like topological insulators and superconductors, could benefit from this approach by providing a mathematical foundation to describe the interplay between field topology and material properties.

Expanding these methods into quantum field theories with higher-dimensional extensions, such as (4+1) or (5+1) dimensional systems, would allow for a more generalized analysis of symmetry breaking, incorporating additional topological invariants and higher-order residue contributions. This broader perspective could uncover hidden symmetries and emergent properties that are not apparent in lower-dimensional analyses. Moreover, the connection between the mathematical techniques discussed and physical observables—such as anomaly coefficients, topological charges, and stability conditions—paves the way for experimental validation of theoretical predictions in both high-energy physics and condensed matter contexts.

Overall, this paper lays the groundwork for a unified framework that bridges mathematical rigor and physical intuition, aiming to deepen our understanding of the profound connections between geometry, physics, and symmetry breaking. Future research will focus on integrating these mathematical tools with numerical simulations and experimental data, enhancing the predictive power of theoretical models in unexplored regimes. As we continue to unravel the complexities of symmetry and topology in high-dimensional field theories, this approach promises to shed light on some of the most fundamental questions in modern physics, from the origins of mass and charge to the behavior of the universe at its most extreme scales. By advancing our mathematical toolkit, we move closer to a more complete theory that encompasses the rich interplay between symmetries and the fabric of reality itself.

\end{document}